\documentclass[usenatbib,useAMS,a4paper]{mn2e}

\usepackage{graphicx}

\usepackage{txfonts}
\def\msun{{\rm M_{\odot}}}

\title [ ]{Ionisation feedback in star formation simulations: The role of diffuse fields}
\author[]{Barbara Ercolano$^{1,2,3}$, Matthias Gritschneder$^{3}$\\
$^1$School of Physics, University of Exeter, Stocker Road, Exeter, EX4 4QL, UK\\
$^2$Universit\"ats-Sternwarte M\"unchen, Scheinerstr. 1, 81679 M\"unchen, Germany\\
$^3$Cluster of Excellence ’Origin and Structure of the Universe’, Boltzmannstr.2, 85748 Garching, Germany\\
$^4$Kavli Institute for Astronomy and Astrophysics, Peking University,Yi He Yuan Lu 5, Hai Dian, 100871 Beijing, China}

\date{Submitted: July 2003}

\pagerange{\pageref{firstpage}--\pageref{lastpage}} \pubyear{2003}

\begin{document}
\def\mnras{MNRAS}
\def\apj{ApJ}
\def\apjl{ApJL}
\def\apjs{ApJS}
\def\araa{ARA\&A}

\def\lta{\mathrel{\spose{\lower 3pt\hbox{$\mathchar"218$}}
     \raise 2.0pt\hbox{$\mathchar"13C$}}}
\def\gta{\mathrel{\spose{\lower 3pt\hbox{$\mathchar"218$}}
     \raise 2.0pt\hbox{$\mathchar"13E$}}}
\def\Msun{{\rm M}_\odot}
\def\msun{{\rm M}_\odot}
\def\Rsun{{\rm R}_\odot}
\def\Lsun{{\rm L}_\odot}
\def\19{GRS~1915+105}
\label{firstpage}
\maketitle

\begin{abstract}

We compare the three-dimensional gas temperature distributions
obtained by a dedicated radiative transfer and photoionisation code,
{\sc MOCASSIN}, against those obtained by the recently-developed Smooth
Particle Hydrodynamics (SPH) plus ionisation code {\sc iVINE} for snapshots
of an hydrodynamical simulation of a turbulent interstellar medium (ISM)
irradiated by a nearby O star. 

Our tests demonstrate that the global ionisation properties of the
region are correctly reproduced by {\sc iVINE}, hence validating
further application of this code to the study of feedback in star
forming regions. However we highlight 
potentially important discrepancies in the detailed temperature distribution. In particular we
show that in the case of highly inhomogenous density
distributions the commonly employed on-the-spot (OTS) approximation yields
unrealistically sharp shadow regions which can affect the dynamical
evolution of the system. 

We implement a simple strategy to include the effects of the diffuse
field in future calculations, which makes use of physically motivated
temperature calibrations of the diffuse-field dominated regions and
can be readily applied to similar codes. We find that 
while the global qualitative behaviour of the system is captured by
simulations with the OTS approximation, the inclusion of
the diffuse field in {\sc iVINE} (called {\sc DiVINE}) results in a
stronger confinement of the cold gas, leading to
denser and less coherent structures. This in turn leads to earlier
triggering of star formation. We confirm that turbulence is being 
driven in simulations that include the diffuse field, but the efficiency is
slightly lower than in simulations that use the OTS approximation. 

\end{abstract}


\section{Introduction}
Ionising radiation from OB stars influences the surrounding
interstellar medium (ISM) on parsec scales. As the gas surrounding a
high mass star is heated, it expands forming an HII region. The
consequence of this expansion is twofold, on the one hand gas is
removed from the centre 
of the potential preventing it from further gravitational collapse. On the
other hand gas is swept up and compressed beyond the ionisation front
producing high density regions that may thus be susceptible to
gravitational collapse (i.e. the ``collect and collapse'' model, Elmegreen
et al 1995). Furthermore, pre-existing, marginally gravitationally
stable, clouds may also be driven to collapse by the advancing
ionisation front (i.e. ``radiation-driven implosion'', Bertoldi 1989,
Kessel-Deynet \& Burkert 2003, Gritschneder et al 2009a). Finally, ionisation feedback is also thought to be a driver for small scale
turbulence in a cloud  (Gritschneder et al 2009b). 
The net effect of photoionisation feedback on the global star
formation efficiency is still, however, under debate.  
\begin{figure*}
\begin{center}
\includegraphics[width=18cm]{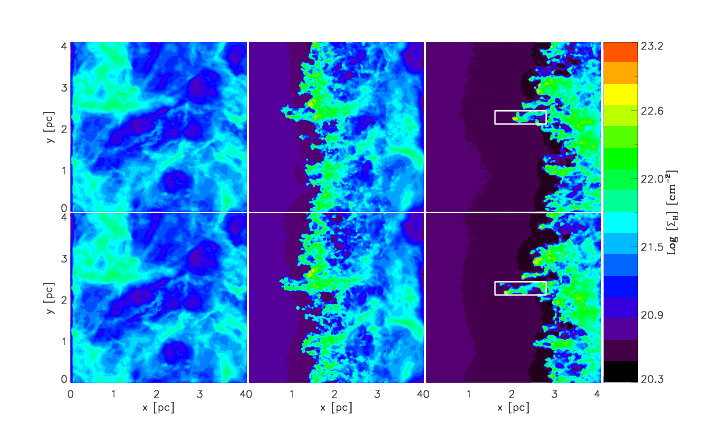}
\caption{Surface density projected in the z-direction of the
  t~=9~kyr (left), 250~kyr (middle) and ~500~kyr (right) snapshots of
  G09b's turbulent ISM simulation. The upper panels show the evolution
  of the gas without diffuse field, the lower panels show the same
  simulation performed with the simplified diffuse field
  implementation discussed in Section~4. The white boundaries on the
  left-hand panels mark the regions compared in Table~1.
The SPH particles were mapped onto a 128$^3$ regular Cartesian grid.} 
\end{center}
\end{figure*}

While the importance of studying the photoionisation process as part
of hydrodynamical star formation simulations has long been widely
recognised, until very recently, due to the complexity and
computational demand of the problem, the evolution of ionised gas
regions had only been studied in rather idealised systems (e.g. Yorke
et al. 1989; Garcia-Segura \& Franco 1996), with simulations often
lacking resolution and dimensions. 
Fortunately, the situation in the latest years has been rapidly improving,
 with more sophisticated implementations of ionised radiation
in grid-based codes presented by (e.g.) Mellema et al (2006), 
  Krumholz et al (2007) and Peters et al (2010). 

Kessel-Deynet \& Burkert (2000) were the first to introduce an
ionisation algorithm into a Smoothed
Particle Hydrodynamical (SPH) code to study radiation-driven
implosion as a possible trigger of star formation. Later, Dale,
Ercolano \& Clarke (2007) presented a much simplified, but fast,
algorithm to consider photoionisation within complex SPH
simulations. When compared to grid
codes, which are based on the solution of the Eulerian form of the
same equations, much higher resolution of very complex flows can be
achieved. Since Dale et al (2007) a number of other ionising radiation
implementations have been developed for SPH codes, including Pawlik \&
Schaye (2008), Altay et al. (2008), Bisbas et al (2009) and very recently Gritschneder et
al (2009a). However high resolution SPH simulations are very
computationally expensive, and even in the current era of parallel
computing, an exact solution of the radiative transfer (RT) and
photoionisation (PI) problem in three dimensions within SPH
calculations is still prohibitive. Necessarily, all the algorithms
mentioned above employ an extremely simplified approach to RT and PI. 
The consequences of such simplification on the conclusions drawn from
the simulations need to be investigated. 

In Dale et al (2007), we performed the only such verification to date against a
fully three-dimensional radiative transfer and 
photoionisation code ({\sc MOCASSIN}, Ercolano et al 2003, 2005 and
2008) for complex density fields obtained from the SPH
calculations. We showed in that case that the agreement on the ionised
fractions in high 
density regions was very good, but low density regions were poorly
represented by the ionisation + SPH code. In this paper we take a
similar approach to test the more recent algorithms developed
in {\sc iVINE} (Gritschneder et al. 2009a).
We find excellent
agreement between the codes for the global ionisation fractions,
however we highlight discrepancies in the temperature
distribution, particularly in shadow regions that are dominated by the
diffuse field, which is not accounted for in iVINE. We test the
consequences of the omission of the diffuse field both on the
hydrodynamical evolution of the structure and on the evolution of the
turbulence spectrum using a simple approach that allows for a more
realistic, but still efficient modelling strategy of the shadow
regions. 

In Section~2 we briefly describe the {\sc iVINE} and {\sc MOCASSIN}
codes and the comparison strategy. In Section~3 we show the results of
this comparison. In Section~4 we discuss a simple approach to
qualitatively include diffuse field effects in {\sc iVINE} and show
the results from these further tests in Section~5. Section~6 contains a brief
summary and future directions. 

\section{Numerical Methods}

\begin{figure*}
\begin{center}
\includegraphics[width=18.0cm]{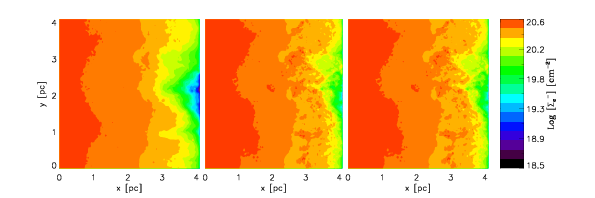}
\caption{Surface density of electrons projected in the z-direction for
  the G09b turbulent ISM simulation at t~=~500~kyr (bottom
  panels of Figure~1). {\it Left:} {\sc iVINE}; {\it Middle:} {\sc MOCASSIN} H-only;
  {\it Right:} {\sc MOCASSIN} nebular abundances.}  
\end{center}
\label{f:sigmane}
\end{figure*}

We have used the {\sc MOCASSIN} code (Ercolano et al 2003, 2005, 2008a) 
to calculate the temperature and ionisation structure of the turbulent
ISM density fields presented by Gritschneder et al (2009b, from now on
G09b) and Gritschneder et al (2010). The SPH quantities were
  obtained with the {\sc iVINE} code 
(Gritschneder et al 2009a) and mapped onto a regular 128$^3$ Cartesian
grid. We briefly describe the set-up of the two codes and summarise
the strategy for our comparative tests. 

\subsection{{\sc iVINE}}
{\sc iVINE} (Gritschneder et al
2009a) is a highly efficient fully parallel implementation of
ionising radiation in the tree-SPH code {\sc VINE} (Wetzstein et al
2009, Nelson et al 2009). While the treatment of hydrodynamics and gravitational forces
is fully three-dimensional the UV-radiation of a massive star is
assumed to impinge along parallel rays onto the simulated domain. To keep the computational
effort small the radiation is assumed to be monochromatic. In
addition, ionising photons reemitted following recombinations are assumed to be
immediately absorbed in the direct neighbourhood (on-the-spot
approximation, e.g. Spitzer 1978). On the
surface of infall the domain is 
decomposed into small rays. Along these rays the radiation is
propagated from SPH-particle to SPH-particle by iterating the
ionisation degree to its equilibrium value. This is done at the
smallest hydrodynamical time-scale, i.e. after each hydrodynamical
time-step. The newly found ionisation degree $\eta$ is then coupled to the
hydrodynamic evolution via the gas pressure of each individual
particle i:
\begin{equation}
P_\mathrm{i} = \left(\frac{T_\mathrm{hot} \eta_\mathrm{i}}{\mu_\mathrm{hot}} +
  \frac{T_\mathrm{cold} (1-\eta_\mathrm{i})}{\mu_\mathrm{cold}}\right)
\frac{k_\mathrm{B} \rho_\mathrm{i}}{m_\mathrm{p}},
\end{equation}
\noindent where $T_{hot}$~=~10kK and $T_{cold}$~=~10K and 
$\mu_\mathrm{hot}=0.5$ and $\mu_\mathrm{cold}=1.0$ are the
temperatures and the mean molecular weights of the ionised and
the un-ionised gas in the case of pure hydrogen,
respectively. $k_\mathrm{B}$ is the Boltzmann constant, 
$m_\mathrm{p}$ is the proton mass
and $\rho_\mathrm{i}$ is the SPH-density of the particle.
 To ensure a correct treatment of the newly ionised particles, their time-step is
reduced according to the ratio of the sound-speeds in the hot and the
cold gas. For a detailed description along with numerical tests see Gritschneder et al
(2009a). The accuracy parameters for the simulations presented here
are as given in G09b.

To map the complex particles distribution obtained with the {\sc
  iVINE} code on a equally spaced Cartesian grid we bin the particles
weighted with the SPH kernel. For each particle the grid cells
which are inside the particles smoothing length are determined. Onto
these cells the mass of the SPH particle is then distributed
according to a weight obtained by equating the kernel for the given
distance between particle and grid cell. All other quantities are
weighted accordingly.

\subsection{{\sc MOCASSIN}}

{\sc MOCASSIN} is a three-dimensional photoionisation and dust
radiative transfer code that employs a Monte Carlo approach to the
frequency resolved transfer of radiation. The code includes all
the dominant microphysical processes that influence the gas ionisation balance
and the thermal balance of dust and gas, including processes that couple the
gas and dust phases. In the case of HII regions ionised by OB stars
the dominant heating process for typical gas abundances is
photoionisation of hydrogen, which is balanced by cooling by collisionally excited
line emission (dominant), recombination line emission, free-bound
and free-free emission. 
The atomic database includes opacity data from Verner et al. (1993)
and Verner \& Yakovlev (1995), energy levels, collision strengths and
transition probabilities from Version 5.2 of the CHIANTI database
(Landi et al. 2006, and references therein) and the improved hydrogen
and helium free-bound continuous emission data of Ercolano \& Storey
(2006). The code was originally developed for the detailed
spectroscopic modelling of ionised gaseous nebulae
(e.g. Ercolano et al 2004, 2007), but is regularly applied to a wide range
of astrophysical environments, including protoplanetary discs
(e.g. Ercolano et al 2008b, 2009, Owen et al 2010, Schisano et al 2010) and
supernova envelopes (e.g. Sugerman et al 2006, Ercolano et al 2007b, Wesson
et al 2009, Andrews et al 2010). 
Arbitrary ionising spectra can be used as well as multiple ionisation
sources whose ionised volumes may or may not overlap, with the overlap
region being self-consistently treated by the code. Arbitrary dust abundances,
compositions and size distributions can be used, with independent grain
temperatures calculated for individual grain sizes. 

In order to compare with {\sc iVINE}, the stellar field was assumed to
be incident along parallel rays, but the subsequent RT was performed in three
dimensions hence allowing for an adequate representation of the
diffuse field. We scaled the incoming stellar field to the value used
by G09b ($F_{Ly}$~=~5$\times$10$^9$ phot/cm$^2$/sec) and
we assumed a blackbody spectrum of 40kK. The exact choice of the
blackbody temperature (within the range of typical effective
temperatures for OB stars) does not have a significant impact for our
application. We did not include dust grains in the calculation in
order to allow a more immediate comparison with  {\sc iVINE}.  

\begin{figure*}
\begin{center}
\includegraphics[width=18.0cm]{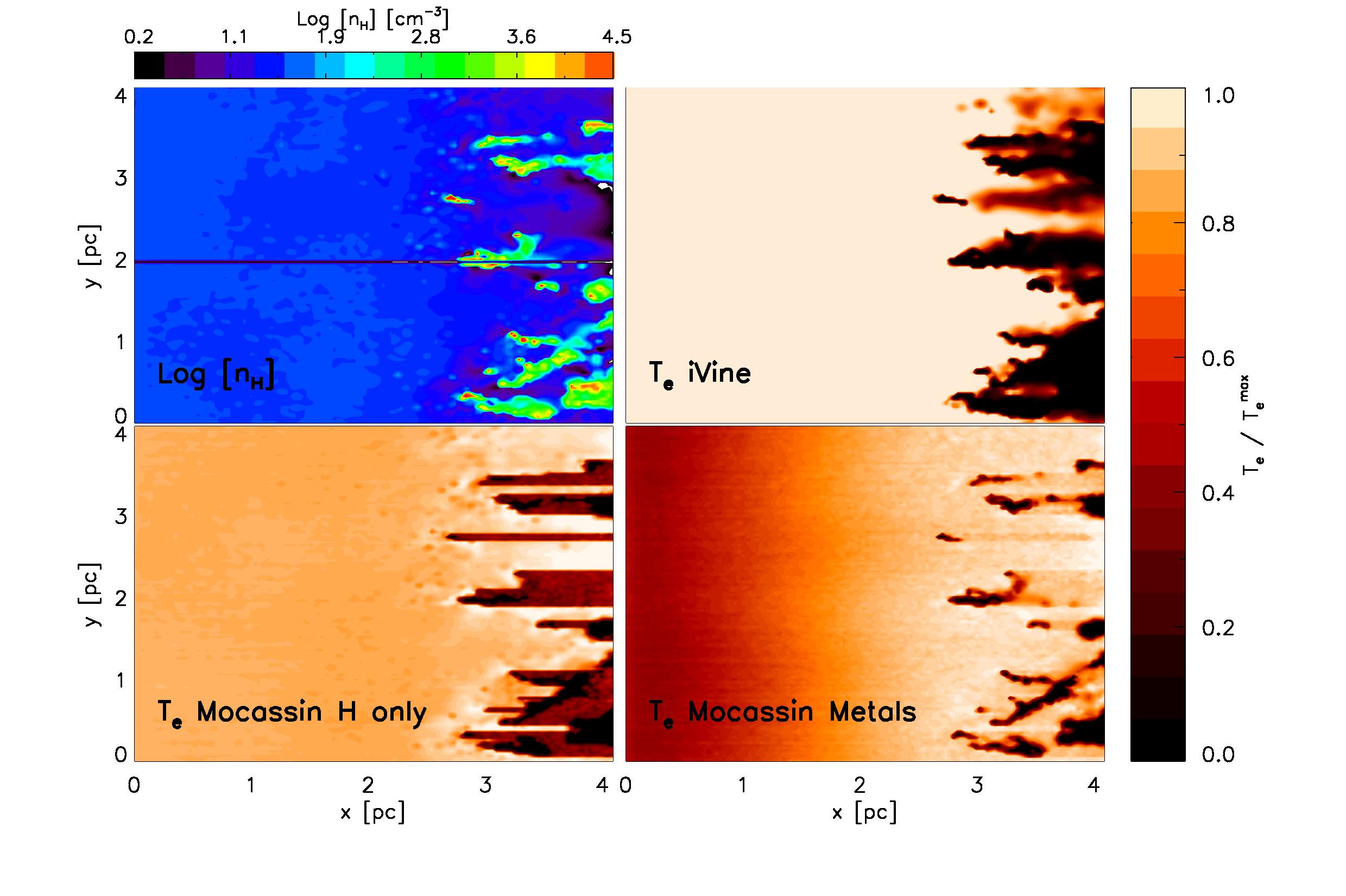}
\caption{Density and temperature maps for the z = 25 slice of the G09b
  turbulent ISM simulation at t~=~500~kyr. {\it Top left:} Gas density
  map; {\it Top right:} electron temperature, $T_e$ as calculated by
  {\sc iVINE};; {\it Bottom left:} electron temperature, $T_e$ as
  calculated by {\sc MOCASSIN} with H-only; {\it Bottom right:}
  electron temperature, $T_e$ as calculated by {\sc MOCASSIN} with
  nebular abundances. The solid black horizontal line shows the
  location chosen for the plots shown in Figure~4.}
\end{center}
\label{f:shade}
\end{figure*}

\subsection{Strategy for comparison test}

The comparisons are based on the simulations of G09b which address the
impact of ionising radiation on a turbulent ISM. These simulations
used  $2\times10^6$ SPH-particles, which are set up to resemble a
medium with an average number density of $300 cm^{-3}$ in a volume of
$(4pc)^3$. To mimic turbulence, a supersonic velocity field with Mach
10 is set up on the largest modes. This setup is allowed to decay
freely down to a turbulent Mach number of 5 before the ionisation
field is switched on. We have taken snapshots of this simulation at
several time-steps and mapped the corresponding SPH particle fields to
a 128$^3$ Cartesian grid.

We have taken snapshots at several time-steps in the turbulent ISM
simulation of G09b and mapped the corresponding SPH particle fields to
a 128$^3$ Cartesian grid. Figure~1 shows the surface density projected
in the z-direction for the t~=~9~kyr (left), t~=~250~kyr (middle), and
t~=~500~kyr (right) snapshots. The top panels show the results from the
original G09b calculation (no diffuse field), while the bottom panels
show the same results for the approximate diffuse field calculation
presented in Section~4. The comparison tests discussed in Section~3
were performed using the 
distributions from the original G09b calculation shown in the top
panels. 

Three-dimensional RT and PI calculations were performed on the
Cartesian grid using {\sc  MOCASSIN} as described above. We run H-only
simulations (referred to 
as ``H-only'') as well as simulations with typical HII region abundances
(referred to as ``Metals''). The elemental abundances for the metal-rich model
were as follows, given as 
number density with respect to Hydrogen: He/H = 0.1, C/H = 2.2e-4, N/H
= 4.0e-5, O/H = 3.3e-4, Ne/H = 5.0e-5, S/H = 9.0e-6. 

We compared the resulting {\sc MOCASSIN} temperature and ionisation
structure grids to {\sc iVINE} with the aim to address the following
questions: (i) Are the global ionisation fractions accurate? (ii) How
accurate is the gas temperature distribution? (iii) What is the effect
of the diffuse field? (iv) How can the algorithm be improved?

\section{Results}
In this section we present the results of our comparison of the
original G09b {\sc iVINE} calculations (top panels in Figure~1) with
{\sc MOCASSIN}.  We limit the discussion to the t~=~500 kyr snapshot,
but we note here that the same conclusions apply to the other timesteps (we
have performed the same tests at several time snapshots). 

\begin{figure}
\begin{center}
\includegraphics[width=8.5cm]{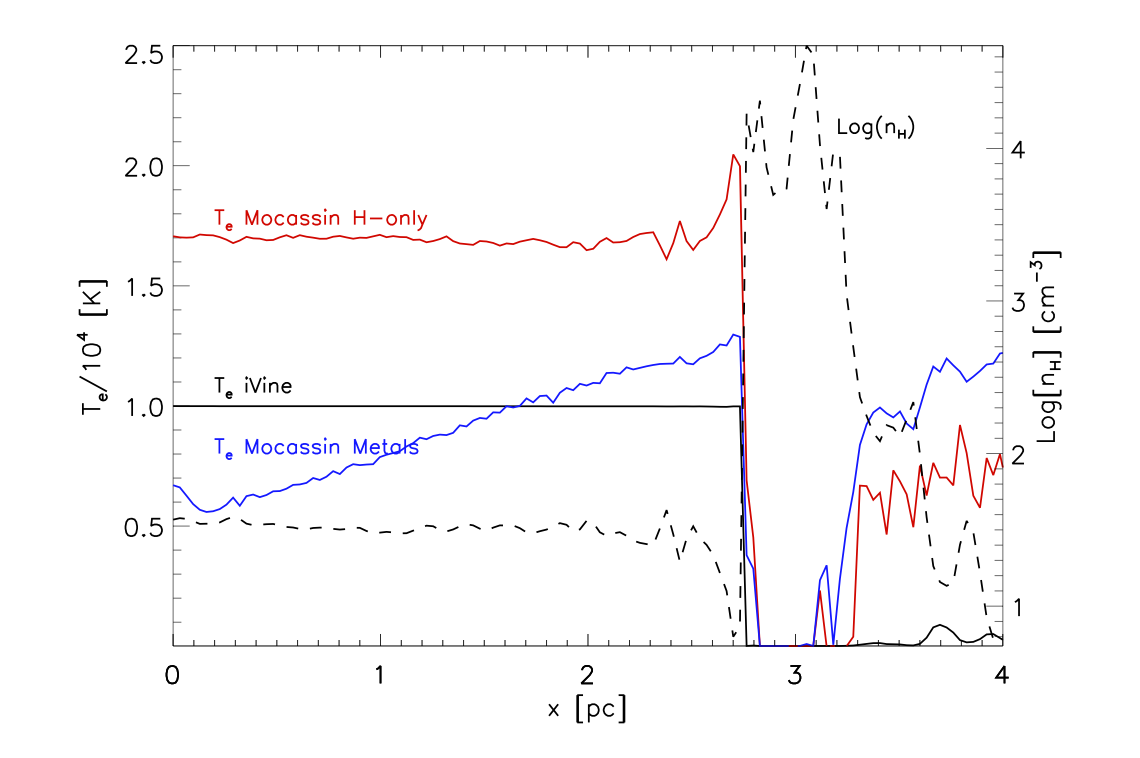}
\caption{Density and temperature profile along a typical ray of the G09b turbulent ISM simulation at t~=~500~kyr. The black dashed line shows the density profile, the black solid line shows the temperature profile calculated by iVINE the solid red and blue profile show the temperature profiles calculated by {\sc MOCASSIN} with H-only and nebular abundances, respectively. }
\end{center}
\end{figure}

\subsection{Global Properties}

Figure~2 shows the surface density of electrons
projected in the z-direction for the G09b turbulent ISM simulation at
t~=~500~kyr. The figure shows that the integrated ionisation
  structure is reasonably well reproduced by {\sc iVINE}. There are
  however some qualitative differences at the rear of the grid
  (X>3~pc) where {\sc iVINE} predicts a lower surface density of
  electrons (i.e. a more neutral gas). This is due to the fact that
  these regions are completely shielded from ionising photons in {\sc
    VINE} by overlapping shadows. In the {\sc mocassin} simulations,
  however, reprocessed photons are able to diffuse amongst the clumps
  and reach the rear of the domain. Nevertheless, these differences
  are small and the global ionisation structure is correctly
determined by {\sc 
  iVINE}, as confirmed by the simple comparison of the total
ionised mass fractions: at t~=~500~kyr, iVINE obtains a total ionised
mass of 13.9\%, while {\sc MOCASSIN} ``H-only'' and ``Metals'' obtain
15.6\% and 14.0\%, respectively. The agreement at other time snapshots
is equally good (e.g. at t~=~250~kyr {\sc iVINE} obtains 9.1\% and {\sc
  MOCASSIN} ``Metals'' 9.5\%). 

It may at first appear curious that the agreement should be better
between {\sc iVINE} and {\sc MOCASSIN} ``Metals'', rather than {\sc
  MOCASSIN} ``H-only'', given that only 
H-ionisation is considered in {\sc iVINE}. This is however simply
explained by the fact that {\sc iVINE} adopts an ``ionised gas
temperature'' ($T_{hot}$) of 10kK, which is close to a {\it typical} HII
region temperature, with {\it typical} gas abundances. The absence of
metals in the {\sc MOCASSIN} ``H-only'' simulations causes the temperature to rise to
typical values close to 17kK, due to the fact that cooling becomes
much less efficient without collisionally excited lines of oxygen,
carbon etc. The hotter temperatures in the ``H-only'' models directly
translate into slower recombinations, as the recombination coefficient
is roughly proportional to the inverse of the temperature. As a
result of slower recombinations the ``H-only'' grids have a slightly
larger ionisation degree. 

\subsection{Ionisation and temperature structure}

\begin{figure}
\begin{center}
\includegraphics[width=8.5cm]{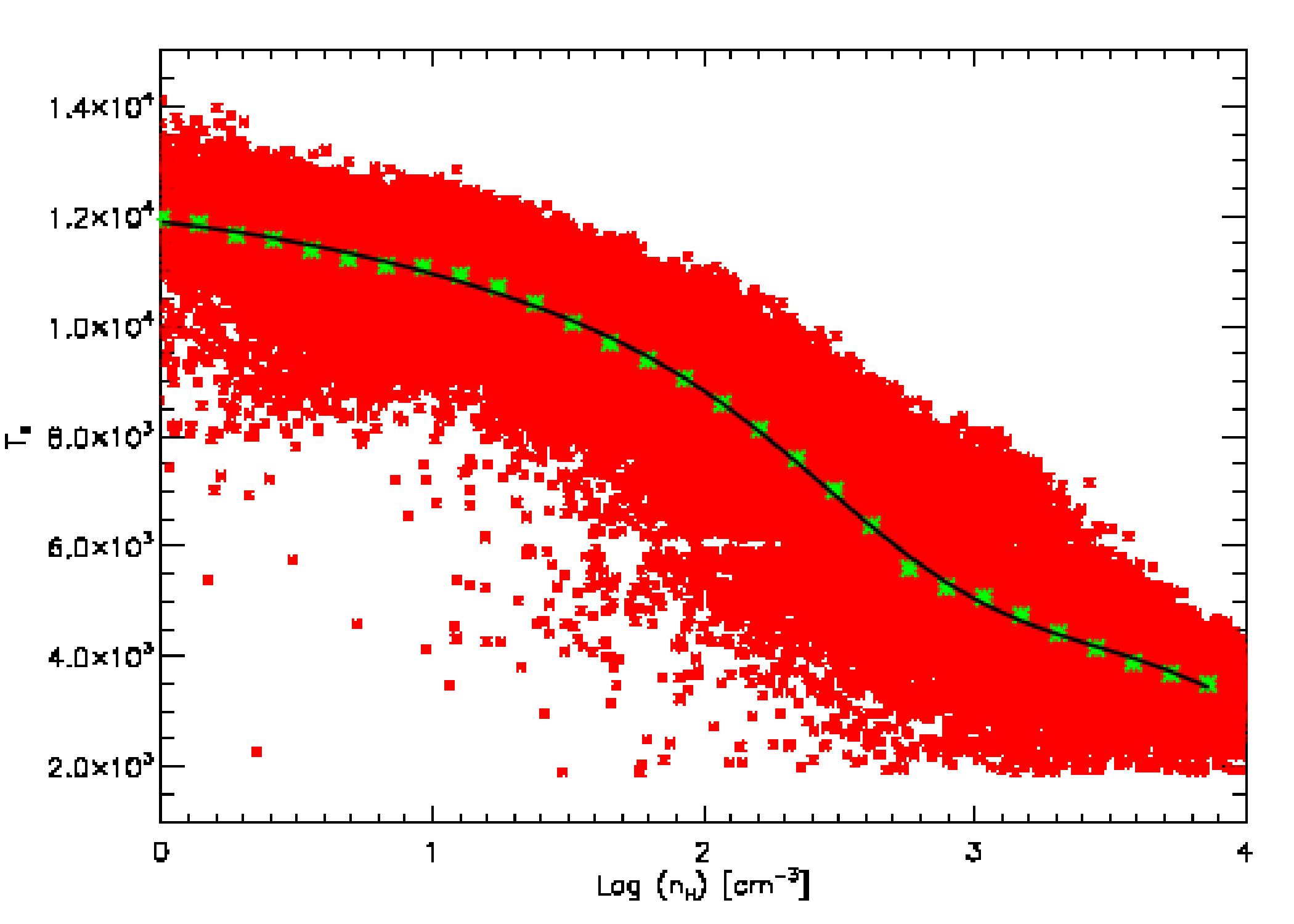}
\caption{Temperature versus density distribution of the shadowed
  cells. The red points are the values for each cell as calculated by
  {\sc MOCASSIN} with nebular abundances; the green points represent
  the binned data and the black solid line represents a Gaussian fit
  to the binned data.}
\end{center}
\label{f:rel}
\end{figure}

The gas temperature of a particle in {\sc iVINE} is calculated
as a simple function of the 
ionisation fraction, $\eta$ i.e. 
\begin{equation}
T_e = T_{hot} \cdot \eta + T_{cold} \cdot (1 - \eta)
\end{equation}
\noindent where $T_{hot}$~=~10kK and $T_{cold}$~=~10K are the
temperatures assigned to fully ionised and neutral material,
respectively.
Accurate gas temperatures are of prime importance as this is how
feedback from ionising radiation impacts on the hydrodynamics of the
system (cf Eq. 1). In Figure~3 we compare the
electron temperatures $T_e$ calculated
by {\sc iVINE} and {\sc MOCASSIN} (``H-only'' and ``Metals'') in a
z-slice of the t~=~500kyr grid.  The top-right panel of the figure
shows the volumetric number density [cm$^{-3}$] map for the selected slice. The large
shadow regions behind the high density clumps in the {\sc iVINE}
calculations are immediately evident in the figure. These shadows are largely reduced in the {\sc
  MOCASSIN} calculations as a result of diffuse field ionisation. The
diffuse field is softer than the stellar field and therefore
temperatures in the shadow regions are lower than in the directly
illuminated regions. Figure~4 shows a more quantitative comparison of the
temperature distribution along a single ray intercepting a high
density clump and its shadow region. The ray is marked in
Figure~3 as the solid black line in the top
left panel. The dashed line in Figure~4 shows the
logarithmic hydrogen number density as a function of distance into the
grid. The solid black solid line is the electron temperature along the
ray as determined by {\sc iVINE}, while the solid red and blue lines
are the electron temperatures determined by {\sc MOCASSIN} ``H-only''
and ``Metals'', respectively. 
The higher temperatures
in the shadow regions of the {\sc MOCASSIN} ``Metals'' model are a
consequence of the Helium Lyman radiation and the heavy elements
free-bound contribution to the diffuse field.
\begin{figure*}
\begin{center}
\includegraphics[width=18cm]{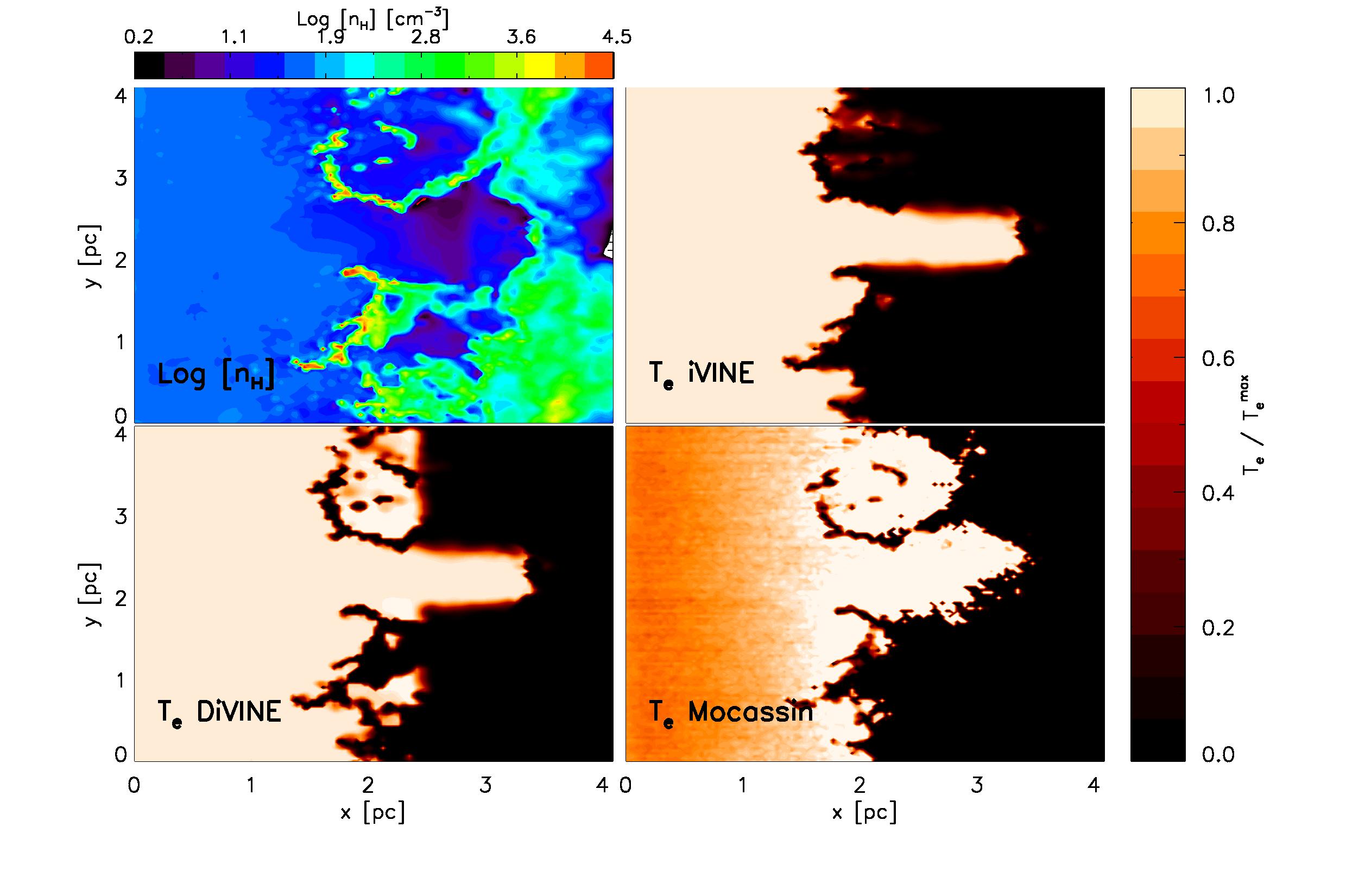}
\caption{Density and temperature maps for the z = 25 slice of the G09b
  turbulent ISM simulation at t~=~250~kyr. {\it Top left:} Gas density
  map; {\it Top right:} electron temperature, $T_e$ as calculated by
  {\sc iVINE};; {\it Bottom left:} electron temperature, $T_e$ as
  calculated by {\sc DiVINE}; {\it Bottom right:}
  electron temperature, $T_e$ as calculated by {\sc MOCASSIN} with
  nebular abundances. }
\end{center}
\end{figure*}

The effects of radiation-hardening and of the recombination of some of
the important cooling ions is also apparent in {\sc   MOCASSIN}'s
temperatures which rise with distance from the star in the directly
illuminated regions. As expected the
effect is much more pronounced in the ``Metals'' model (a steeper
temperature distribution in metal-rich regions is a well known effect,
e.g. Stasinska~2005, Ercolano et al 2007). Radiation
hardening is accounted for in some ``monochromatic'' RT codes, by 
pre-calculating the effective photoionisation x-section as a function of
the Lyman-limit optical depth by integrating over the extinguished
stellar spectrum, and doing the same for the photoelectric heating
integral (e.g. {\sc C2Ray}, Mellema et al. 2006). Helium, heavy elements or
dust effects are not included, by this treatment however and most
importantly the effects of different coolants acting in different
regions (which is dominating the temperature distribution in the
directly illuminated regions of the {\sc MOCASSIN} Metals
calculations) cannot be accounted for in this way. In future work we
plan to provide simple temperature calibrations to account for
the environmentally driven temperature gradients in the directly
illuminated regions. We note however that errors in the temperature of
the directly illuminated regions are unlikely to exceed factors of two
at most, the errors on the temperature of the diffuse field dominated
regions, on the other hand, are typically three orders of magnitude
(10K compared to 10kK). Thus, in the remainder of this paper we shall
focus on the larger temperature differences in the shadowed
  regions and how they may affect
the hydrodynamical evolution of the system. 

\section{Gas temperatures in shadow regions}

\begin{figure*}
\begin{center}
\includegraphics[width=18cm]{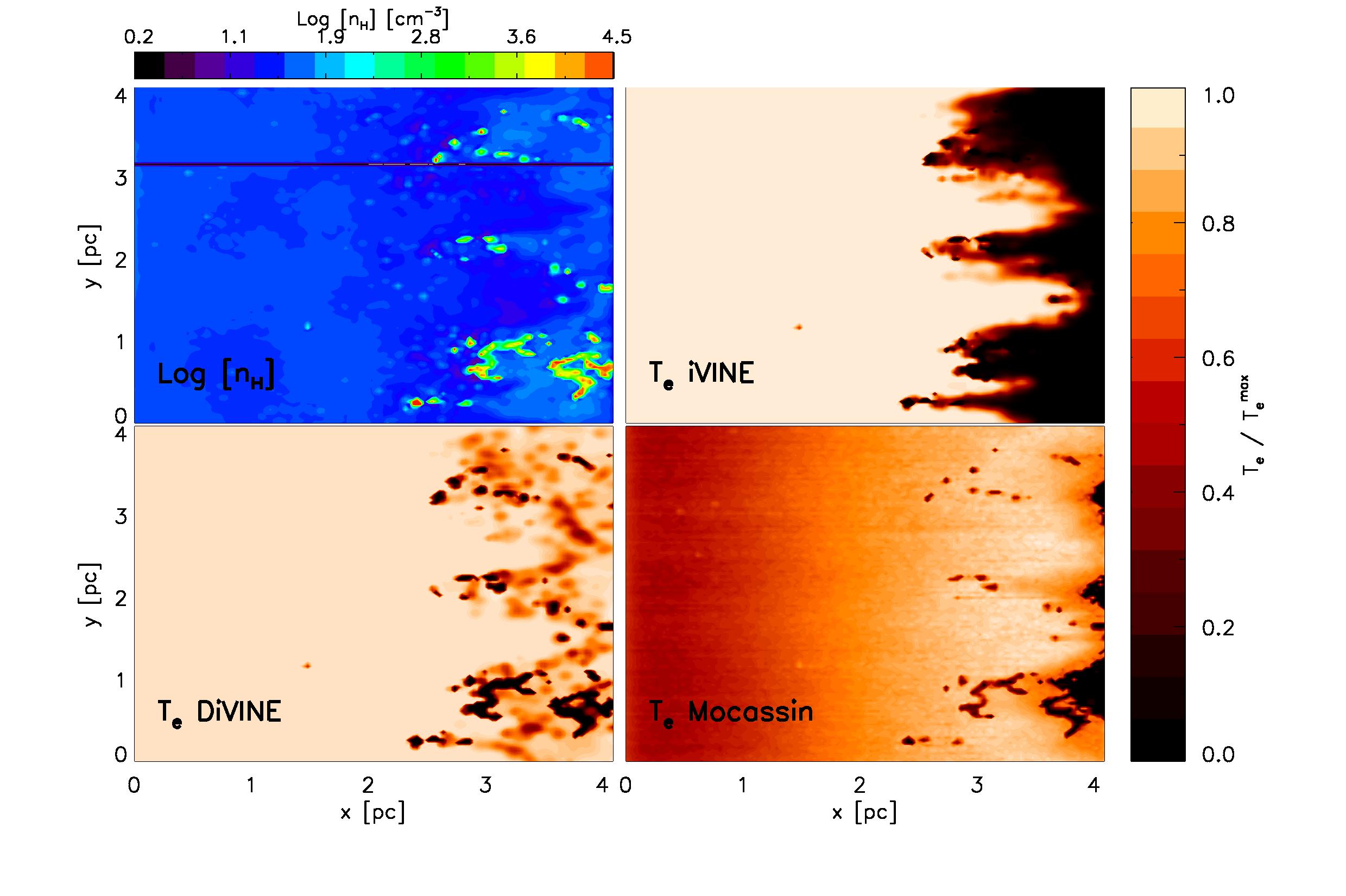}
\caption{Density and temperature maps for the z = 25 slice of the G09b
  turbulent ISM simulation at t~=~500~kyr. {\it Top left:} Gas density
  map; {\it Top right:} electron temperature, $T_e$ as calculated by
  {\sc iVINE};; {\it Bottom left:} electron temperature, $T_e$ as
  calculated by {\sc DiVINE}; {\it Bottom right:}
  electron temperature, $T_e$ as calculated by {\sc MOCASSIN} with
  nebular abundances. }
\end{center}
\end{figure*}

As {\sc iVINE} solves the transfer along plane parallel rays using the
on-the-spot approximation (e.g. Spitzer 1978, Osterbrock \& Ferland 2006, page 26), it has
currently no means of bringing ionisation (and hence heating) to
regions that lie behind high density clumps. This creates sharp
shadows with a large
temperature (pressure) gradient between neighbouring direct and
diffuse-field dominated regions, which may have important implications
for the dynamics, particularly with respect to turbulence
calculations. 

In order to explore the significance of the error introduced by the
OTS approximation on the dynamics of the system, we propose here a simple
strategy to include the effects of the diffuse 
field in {\sc iVINE}, which can be readily extended to other similar
codes. It can be summarised in the following steps: (i) identify the
diffuse field dominated regions (shadow); (ii) study the realistic
temperature distribution in the shadow region using fully frequency
resolved three-dimensional photoionisation calculations performed with
{\sc MOCASSIN} and parameterise the gas temperature in the shadow
regions as a function of (e.g.) gas density; (iii) implement the
temperature parameterisation in {\sc iVINE} and update the gas
temperatures in the shadow regions at every dynamical time step
accordingly. 

\begin{figure*}
\begin{center}
\includegraphics[width=17cm]{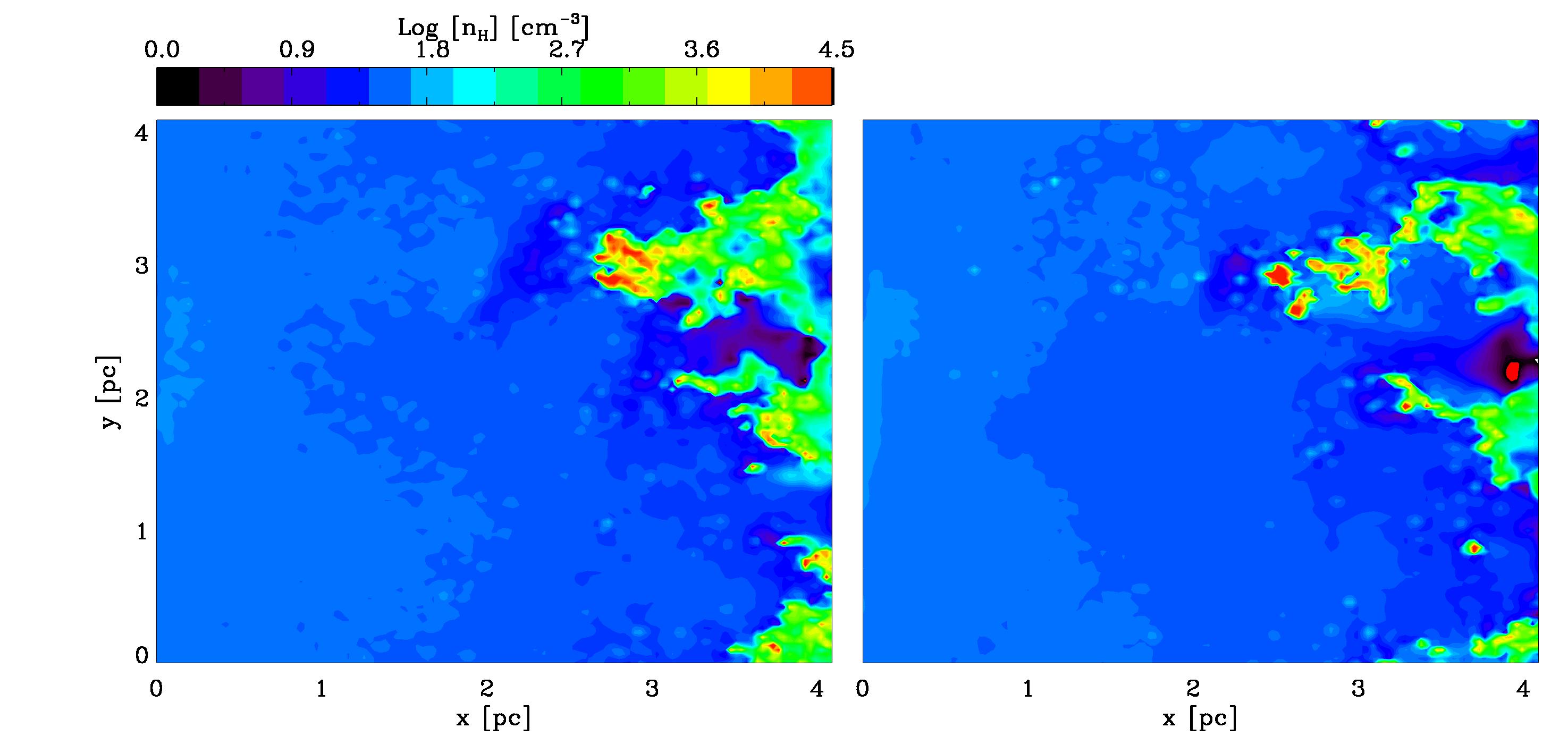}
\caption{Volume density slices at t = 500kyr for the {\sc iVINE}
  (left) and {\sc DiVINE} (right) runs.}
\end{center}
\end{figure*}

In what follows we describe our zeroth order implementation of the
above strategy for the case at hand, but we stress that this can be
adapted to different situations. In future work we will present a
generalised version of the procedure detailed below, which will
include temperature parameterisations for both the direct and
diffuse-field dominated regions and will account for different
environments and geometrical setups of the codes. This requires a
parameter space study and is beyond the scope of the current work,
where we only aim to assess the relevance of relaxing the OTS
approximation for highly inhomogenous cases. 

\subsection{Step i: Identify the shadows} 
In the case of {\sc iVINE}, we identify the
diffuse-field dominated regions by means of simple criteria that need 
only be evaluated along a ray, such that no significant overheads are
introduced by this step. A particle / grid-cell is initially defined to be
'shadowed' if {\sc iVINE} would assign a temperature of $T<100$~K
  to it, i.e. if its ionisation degree is $\eta<10^{-2}$.

\subsection{Step ii: Study the temperature structure in the shadows}
We study the {\sc MOCASSIN}-determined temperature structure of 
the shadow regions identified in Step i, for several time snapshots
in the G09b simulation. For each snapshot we plot the gas temperature
against density for all the shadowed cells in the entire
grid. Figure~5 shows the results for the t~=~500~kyr 
snapshot. The red marks represent each individual grid cell in the
{\sc MOCASSIN} grid, while the yellow asterisks represent the averaged value, which is then fit by the curve shown by the solid line. 
The fit was obtained by a six term Gaussian fit of the form:
\begin{equation}
y = A_0*exp(-(z^2)/2) + A_3 + A_4 x + A_5 x^2
\label{e:rel}
\end{equation}
\noindent where x = $T_e$, y = $Log(n_H)$, z = (x - A$_1$)/A$_2$ . The fit coefficients for the
curve shown are A$_0$= -4378.10, A$_1$= -0.0275, A$_2$= 0.616, A$_3$= 17715.8, A$_4$
= -5322.37, A$_5$ = 395.13. The form of this relationship does not vary much for different
time snapshots and it is mainly controlled by the shape of the direct
field and the metallicity of the region. 

The near-invariance of the T$_e$-n$_H$ relation
at different times is certainly encouraging, however a number of
caveats need also be noted here. Most importantly the fact that not
all of the shadow cells identified in Step~i obey Eqn~3. A
large fraction of these cells are in fact truly shadowed and have
temperatures of $<$100~K also according to {\sc MOCASSIN} (these
cells lie below the lower y-axis cutoff in Figure~5). 
These are cells that are located in regions where the diffuse photons
cannot penetrate, thus, in this 
respect, applying our simple shadow region criteria indiscriminately to
{\it all} gas would severely overestimate the thermal effects of the diffuse
field. We thus apply a further density criterion that only allows low
density cells ($n_H < 100 cm^{-3}$) to be heated by the
diffuse field. This is justified by 
the fact that our grids show that the number of cells that follow
Equation~3 varies according to the density of the cell. At t~=~500~kyr
we find that over 90\% of cells with n$_H < 10 cm^{-3}$, obey Equation~2, this
goes down to 75\% for cells with $10 cm^{-3}< $ n$_H < 100 cm^{-3}$
and then down to less than 10\% for cells with n$_H > 100 cm^{-3}$. 
While this density cut-off will still slightly overestimate the effect of the
diffuse field in the $10 cm^{-3}< $ n$_H < 100 cm^{-3}$ regime, we
note that a temperature of 10~K is only really expected for
dark cloud conditions, perhaps met only in the higher densities
regions in the simulations. The rest
of the gas at lower densities in the true shadow zone (i.e. those
cells that do not obey Equation~2) is more
realistically described by photon dominated region (PDR)
conditions, with temperatures closer to 1000~K, rather than
10~K. So the error introduced by applying Equation~2 to the true
shadow zone in the $10 cm^{-3}< $ n$_H < 100 cm^{-3}$ range is still
smaller than if those cells had been assigned their original
temperature of 10~K.  

A second order problem is posed by the large scatter of the data (red
points) shown in Figure~5. Typical errors of $\sim$50\% in
temperature are to be expected with this method.The temperature
  at a given location in an ionised region is expected to roughly
  scale with the local ionisation parameter, which is defined as the
  ratio of the ionising flux to the gas density at that location. A
  temperature-density relation as the one shown in Figure 5 thus
  ignores the local variations of the diffuse field intensity and
  shape. The large scatter shown in Figure 5 is therefore a
  shortcoming of this method, but has the advantage of avoiding the
  large computational overheads of a full diffuse field
  calculation. We stress, however, that the effect of these errors on the
dynamics is not large. Again, the error introduced by assuming the particle to be
in the cold phase due to the lack of diffuse radiation is much larger
than the spread in this relation. 
 
\subsection{Step iii: Implement temperature corrections}
Once a simple description of the temperature distribution in the
shadow region is available from Step~ii, this can be implemented as a
correction to the SPH temperatures and thus pressure. In the case of the current {\sc
  iVINE} implementation this simply requires that at every dynamical
time step, all particles that fulfil the criteria defined in Steps~i
and ii be identified and their temperature re-assigned according to the
parameterised curve obtained in Step~ii. To avoid unphysical
  heating far beyond the ionisation front, we impose an additional
  constraint. A particle only gets heated if the total
  ionised mass inside the current slab perpendicular to the in-falling
  radiation is smaller then the diffusively heated mass inside this slab. Note
  that this constraint implies that in the last diffusively heated
  slab all radiation emitted by recombinations is only absorbed in the
  adjacent cold structures and not in the hot surrounding gas. This
  therefore represents the case for maximum efficiency for the diffuse
  radiation, but, as demonstrated in Section~5.1, we find that no
  significant overheating of the shadows is introduced  by our
  procedure. 

We note that this approach in principle allows for environmental variables, such as
the hardness of the stellar field and the metallicity and dust content
of the gas to be accounted for in the SPH calculation, since their
effect on the temperature distribution is folded in the
parameterisation obtained with {\sc MOCASSIN}. In most cases however,
environmental effects on the temperature may be smaller than
the scatter showed by the temperature points in Figure~5. 

In the simple implementation of the method presented here, we apply a
density cut to the particles to be heated by the diffuse field, which
allows us to use a rough but fast criterion for the identification of
the diffuse field dominated particles without 
over-estimating the effects of the diffuse field. As shown in Figure
5, the temperature only varies by approximately 50\% for number
densities lower than 100~cm$^{-3}$, meaning that the qualitative
behaviour of the system would not change significantly if an average
value for the temperature were to be used in place of Equation~3. The
application of Equation~3, however, does not add a significant
computational overhead and it is therefore preferred here in view of
future calculations where we aim at refining the method to allow the
application of a temperature correction to the whole density range,
where the temperature variation is larger and the application of a
single temperature value would yield larger errors.

\section{Diffuse field effects on hydrodynamics}

We have compared the results of the original {\sc iVINE} run with
those obtained after implementing the diffuse field strategy outlined
in the previous section. We will refer to the diffuse field
implementation of {\sc iVINE} as {\sc DiVINE} (Diffuse field {\sc
  iVINE}). 

\subsection{Temperature structure in the shadows}
As an a-posteriori check of the diffuse field implementation in {\sc
  DiVINE},
 we compare the gas temperatures at individual slices in the simulation at times
250 and 500~kyr in Figures~6 and 7, respectively. The top right and bottom left and
right panels show the gas temperatures calculated using {\sc iVINE},
{\sc DiVINE} and {\sc MOCASSIN}, respectively, for the density field
shown in the top left panels. We find that, while there are still some
obvious differences between the {\sc DiVINE} and {\sc MOCASSIN}
results, the improvement over the original {\sc iVINE} runs is
certainly encouraging, particularly when considering the minimal
computational overhead introduced by this procedure.  We also note
that the diffuse field heated mass in {\sc DiVINE} and {\sc MOCASSIN}
are very similar. 

Some noticeable differences can be seen in Figures~6 and 7 include a
hot region predicted by {\sc DiVINE} in Figure~6 approximately at
(x,y) = (2.2, 0.7) pc which is not present in {'\sc mocassin} and
cooler regions at large x predicted by {\ mocassin} in Figure~7. These
are both examples in which material in the true shadow has been
erroneously been heated by our diffuse field algorithm in {\sc
  DiVINE}. This only represents a small quantity of gas and has no
significant effect on the dynamical evolution of the system. 

\subsection{Clumps, Pillars and Horse-heads}

\begin{figure}
\begin{center}
\includegraphics[width=8.5cm]{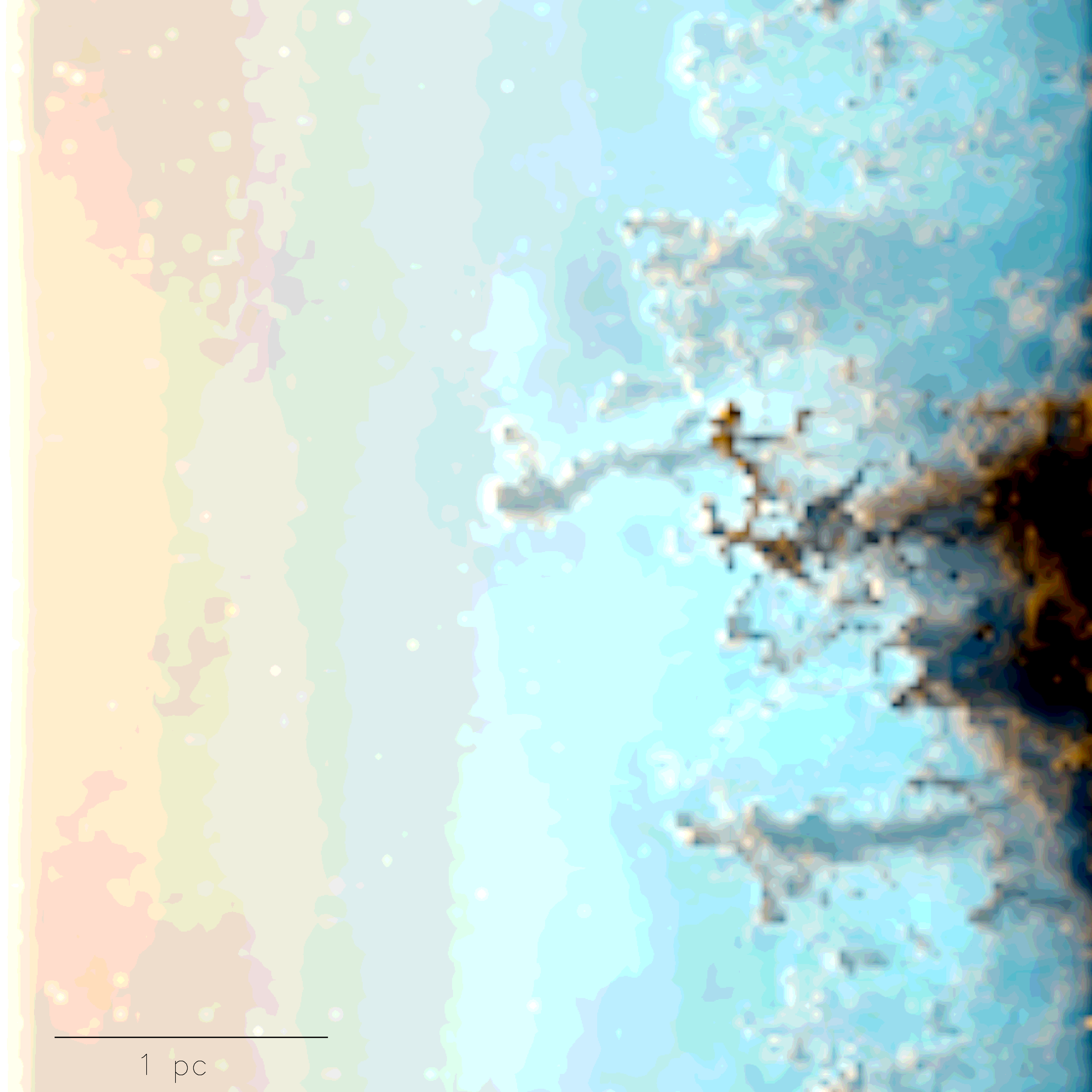}
\caption{False colour composite image of the {\sc DiVINE} simulation at
  t = 500kyr. Red = H$\alpha$, blue =
  $[$OIII$]\lambda\lambda$5007,4959 and green = combination.}
\end{center}
\end{figure}

Figure~1 shows the evolution of the surface density
distribution for the standard {\sc iVINE} run (upper panels) and the
{\sc DiVINE} run (lower panels). The figure suggests that the heating of low density
gas by the diffuse field in the shadow regions results in a clumpier
medium at late times. Comparison of volume density slices, like those
shown at t = 500kyr in Figure~8, further highlight that a clear
effect of the inclusion of the diffuse field is that the pillars are
eroded from the back, promoting the detachment of clumps. This effect
is less visible in integrated surface density maps as those shown in
Figure~1. 

\begin{table}
\begin{center}
\begin{tabular}{lccccc}
\hline
Sim & $M$ &  $\bar{\rho}_\mathrm{p}$ & 
$\bar{\Sigma}$ &
$\sigma$ &  
$\bar{v}_\mathrm{x}$ \\
 & $[\msun]$ &  $[{\thinspace m_\mathrm{p}\thinspace\mathrm{cm}^{-3}}]$ & 
$[\mathrm{g cm}^{-2}]$ & \, $ [{\thinspace\mathrm{km}\thinspace\mathrm{s}^{-1}}] $ \, &  
\, $ [{\thinspace\mathrm{km}\thinspace\mathrm{s}^{-1}}] $ \\
G09b & $12.6$ & $4.56\times 10^{4}$  & $1.52\times 10^{-3}$ & $1.1\pm0.7$ &  $4.8\pm0.9$ \\
DiVINE & $12.0$ & $6.33\times 10^{4}$  & $1.89\times 10^{-3}$ & $0.9\pm0.5$ &  $4.2\pm0.8$ \\
\hline
\end{tabular}
\caption{Comparison of {\sc DiVINE} and {\sc iVINE} at t =500 kyr. Listed are the
mass, mean density, mean surface density, velocity dispersion and the x-velocity
away from the source of the most prominent structure. As the structure
is not homogeneous we also include the standard deviation across the
structure for both velocities. \label{TAB_compare}}
\end{center}
\end{table}

The detailed comparison of the main structure obtained with {\sc
  DiVINE} to the one obtained in G09b is given in Table 1. The regions
compared are marked by the white boundaries drawn on the left hand
panels of Figure~1. It is
interesting to see that the total mass is not strongly
affected. However, the {\sc DiVINE} structure is denser and
spatially thinner. The higher density leads to a slower motion away
from the ionising source. By tracing the particles backwards in
time we are able to determine that 75\% of the material of the pillar
in G09b ends up inside the pillar in  {\sc DiVINE} as well.
Another effect of the higher density is enhanced triggered star
formation. At $t=500$~kyr the star formation is still
identical to G09b. The first star forms in the tip of the secondary pillar, at the same
location and with similar properties as before. After that the star
formation is different. A second star forms in {\sc DiVINE}
at the tip of the primary pillar after $t\approx525$~kyr,
while it forms much later in the G09b simulation (at $t\approx735$~kyr). 

Altogether the overall evolution is similar and the formation
of structures is still governed by the initial turbulence, especially
in high density regions, which are only weakly effected by diffuse
radiation. However, small-scale dynamics and especially the triggered
star formation change due to the inclusion of diffuse radiation.

We have also considered the observational appearance of such clumpy structures in
typical narrow band filters, by computing the H$\alpha$ and 
$[$OIII$]\lambda\lambda$5007,4659 line
emission and producing a false colours image along a line of sight
parallel to the z-axis. We used the interstellar extinction curve of
Weingartner \& Draine (2001) for a Milky Way size distribution for
R$_V$=3.1, with C/H = b$_C$ = 60 ppm in log-normal size distributions, but
renormalised by a factor 0.93. This grain model is considered to be
appropriate for the typical diffuse HI cloud in the Milky Way. The
result is shown in Figure~9, which shows 'pillar-like' as well as
'horse-head-like' structures very much resembling the famous Hubble
Space Telescope images. Hence, even though, the structures are mostly
detached from the parent cloud, the appearance is still that of
coherent pillars. 

\subsection{Photoionisation-driven Turbulence}

G09b found that turbulence on small scales could be sustained by the
ionising radiation. We have revisited this question using the results
obtained from {\sc DiVINE} and we confirm the general conclusion that
turbulence is driven by ionising radiation. A comparison of the
{\sc iVINE} and {\sc DiVINE} turbulence spectra can be seen in
Figure~10, where we plot the
specific energy as a function of wave number (large wave numbers =
small spatial scale) for the original control run with no ionising
radiation presented by G09b (dotted lines), G09b's standard {\sc iVINE} run
(dashed lines) and the {\sc DiVINE} run presented in this work (solid
lines).  The efficiency of the
driving is slightly reduced in the {\sc DiVINE} calculations. We
ascribe this effect to the higher compression of the cold gas due to
the existence of the diffuse phase. The cold gas therefore experiences
more shocks initially and is constrained more tightly, leading to lower
turbulent motions.
\begin{figure}
\begin{center}
\includegraphics[width=9cm]{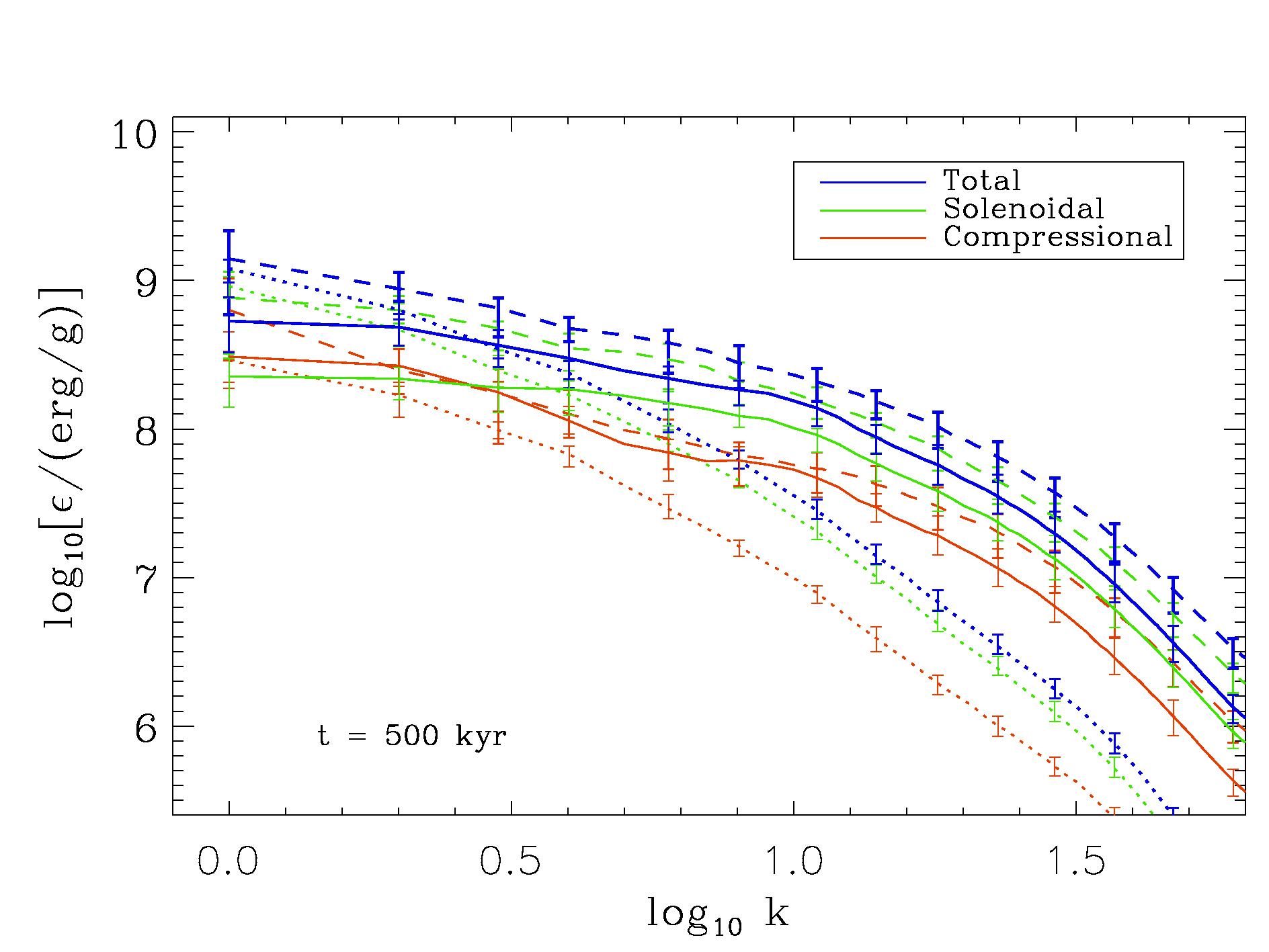}
\caption{Specific energy as a function of wave number. Dotted lines:
  {\sc VINE} run with no ionisation; solid lines: {\sc iVINE} run;
  dashed lines  {\sc DiVINE} run} 
\end{center}
\end{figure}

\section{Conclusions}
 
We have presented a detailed comparison of the ionisation and
temperature structure for a turbulent ISM simulation performed with
the SPH+ionisation code {\sc iVINE} (Gritschneder et al 2009a,b, 2010)
against the solution obtained with the photoionisation
code {\sc MOCASSIN} for snapshots of the density distribution.  {\sc iVINE} treats hydrodynamics, gravitational forces and
  ionisation simultaneously. The ionisation is calculated by making
  use of the  'on-the-spot' (OTS) approximation.
The {\sc MOCASSIN} code (Ercolano et al 2003, 2005, 2008a) is fully
three-dimensional and includes an exact treatment for the frequency resolved transfer of
both the stellar (direct) and diffuse radiation
fields. {\sc MOCASSIN} includes all the microphysical processes that
dominate the thermal and ionisation balance of the ionised gas,
providing realistic temperature and ionisation distributions. 

Our tests show that {\sc iVINE} and {\sc MOCASSIN} agree very well on
the global properties of the region (i.e. total ionised mass
fraction and location of the main ionisation front), but we note discrepancies in the temperature
structure, particularly in the shadow regions. These tend to be cold
and neutral in the {\sc iVINE} plane-parallel stellar-field-only prescription
, while {\sc MOCASSIN} obtains a range of ionisation levels
and temperatures that can be very crudely described as a function of
density. 

We have developed a computationally inexpensive strategy to include
the thermal effects of the diffuse field, as well as accounting for
environmental variables, such as gas metallicity and stellar spectra
hardness. The method relies on the identification of the shadow region
via simple criteria and application of a temperature parameterisation
that was obtained a-priori using {\sc MOCASSIN}. The method can be
readily extended to other hydrodynamical codes (both SPH and
grid-based). 

We evaluate the effects of diffuse fields by comparing runs with the
standard {\sc iVINE} and the diffuse field 
implementation {\sc DiVINE}. In agreement with previous studies (Raga
et al 2009), we find that the overall qualitative behaviour of the
system (i.e. the formation of what appear to be pillar like
structures) is similar in the two runs. Nevertheless our models demonstrate
that the diffuse field has important quantitative effects
on the hydrodynamical evolution of the irradiated ISM. In particular
we note that {\sc DiVINE} predicts denser and less coherent structures
which are much less attached to 
the parental cloud. This is due to the higher compression of the
cold structures by the diffusively heated material inside the pillars
trunks. Triggered star formation is promoted by this
effect to a much earlier time.
The compression also
affects the turbulence spectrum of the system. We confirm the driving of turbulence
by the ionising radiation, but with a slightly reduced efficiency
compared to previous calculations with {\sc iVINE} using the OTS approximation. 

\section{Acknowledgments}
We thank Will Henney and Nate Bastian for useful discussion. We thank
the referee for a constructive report that helped us improve the
clarity of our paper. MG acknowledges additional
funding by the China National Postdoc Fund Grant No. 20100470108 and
the National Science Foundation of China Grant No. 11003001.  
Calculations presented here were partially using
the University of Exeter Supercomputer, the KIAA Computational Cluster
and the University of Munich SGI Altix 3700 Bx2 supercomputer
that was partly funded by the DFG cluster of excellence 'Origin and Structure of the Universe'.

\end{document}